\documentclass[preprint]{aastex61}

\usepackage{physics}
\usepackage{savesym}
\savesymbol{tablenum}
\usepackage{siunitx}
\restoresymbol{SIX}{tablenum}

\submitjournal{PASP}

\shorttitle{Imaging Polarimetry of the 2017 Solar Eclipse}
\shortauthors{Vorobiev et al.}

\begin{document}

\title{Imaging Polarimetry of the 2017 Solar Eclipse with the RIT Polarization Imaging Camera}

\correspondingauthor{Dmitry Vorobiev}
\email{dmitry.vorobiev@lasp.colorado.edu}

\author{Dmitry Vorobiev}
\affil{Laboratory For Atmospheric and Space Physics \\
University of Colorado \\
1234 Innovation Drive \\
Boulder, CO 80303, USA}

\author{Zoran Ninkov}
\affil{Center for Imaging Science \\
Rochester Institute of Technology \\
54 Lomb Memorial Drive \\
Rochester, NY 14623, USA}

\author{Lee Bernard}
\affil{University of California, Santa Cruz \\
1156 High St \\
Santa Cruz, CA 95064, USA}

\author{Neal Brock}
\affil{4D Technology Corporation \\
3280 E. Hemisphere Loop, Suite 146 \\
Tucson, AZ 85706, United States}



\begin{abstract}
{\small In the last decade, imaging polarimeters based on micropolarizer arrays have been developed for use in terrestrial remote sensing and metrology applications. Micropolarizer-based sensors are dramatically smaller and more mechanically robust than other polarimeters with similar spectral response and snapshot capability. To determine the suitability of these new polarimeters for astronomical applications, we developed the RIT Polarization Imaging Camera to investigate the performance of these devices, with a special attention to the low signal-to-noise regime. We characterized the device performance in the lab, by determining the relative throughput, efficiency, and orientation of every pixel, as a function of wavelength. Using the resulting pixel response model, we developed demodulation procedures for aperture photometry and imaging polarimetry observing modes. We found that, using the current calibration, RITPIC is capable of detecting polarization signals as small as $\sim0.3\%$. To demonstrate the stability of RITPIC's calibration and its extreme portability, we performed imaging polarimetry of the Solar corona in Madras, Oregon during the total Solar eclipse of 2017. The maximum polarization we measured was $\sim46\%$, which agrees well with the maximum value predicted for a Thomson scattering corona. Similarly, we found no strong deviations in the angle of linear polarization from the tangential direction. The relative ease of data collection, calibration, and analysis provided by these sensors suggest than they may become an important tool for a number of astronomical targets.}
\end{abstract}




\section{Introduction} \label{sec:introduction}

A total Solar eclipse is an awesome spectacle, which offers a rare opportunity for scientific observations that cannot be made otherwise. The most striking feature of a total Solar eclipse is the appearance of the Solar corona, which is otherwise too faint to be seen through the foreground of the daylight sky. Although the invention of the coronagraph allowed much of the corona to be observed without the aid of the Moon, the sky foreground remains a formidable adversary. Even the coronagraphs of the Large Angle Spectroscopic Coronagraph (LASCO) instrument on the space-based Solar and Heliospheric Observatory (SOHO) are restricted by diffraction and scattering effects to coronal regions further than 1.1 R$_{\Sun}$ from the Solar center \citep{Brueckner1995}. A total eclipse provides access to coronal regions that cannot otherwise be observed. 

Of particular interest, is the polarization of the corona. Near the Sun ($<$ 10 $_R{\Sun}$), the coronal brightness can be attributed to Thomson scattering of chromospheric photons off the free electrons of the K corona and the more complex scattering off dust particles that make up the F corona. Although the F corona only becomes substantial compared to the K corona at a radial distance of $\geq$ 5 R$_{\Sun}$, it has been demonstrated by \cite{vanDeHulst1950} that along the line of sight, light from the F corona dilutes the high intrinsic polarization ($60\% - 70\%$) of the K corona, resulting in observed peak values of $40\% - 50\%$ (in the equatorial regions). Despite the difficulty of separating the contributions from these two components, the polarized brightness of the corona offers a unique probe of the electron density of the K corona \citep{Minnaert1930, vanDeHulst1950, Quemerais2002}.

The path of a Solar eclipse does not usually coincide with the location of an established observatory, which requires the motivated observer to travel with her equipment. In the past, eclipse observations have been made with the naked eye (for example, \cite{Arago1854}), photographic film \citep{Koutchmy1971, Koutchmy1993, Kim2017}, and, more recently, electronic sensors \citep{Skomorovsky2012}. The non-linear response of photographic film offers a large dynamic range, which is critical to observations of the corona, whose brightness changes by a factor of $>$ 1000 over a distance of just a few Solar radii. However, that same non-linearity is detrimental to polarimetric efforts, which rely on accurate, quantitative comparisons of brightness obtained through several polarimetric channels. 

Polarization-sensitive imaging arrays \citep{Nordin1999, Brock2011, Vorobiev2018} are an attractive platform for eclipse polarimetry, due to their small size and ease of use. In this work, we present our efforts perform broadband imaging polarimetry of the Solar corona, during the total solar eclipse of August 21, 2017. We present our calibration approach and our measurements of the degree and angle of linear polarization, with an emphasis on overall polarimetric sensitivity (\textit{i.e.}, ``uncertainty''). 

\section{Polarization-Sensitive Imaging Arrays} \label{sec:ritpicDescription}

\begin{figure}[ht]
\centering
\includegraphics[width=\textwidth]{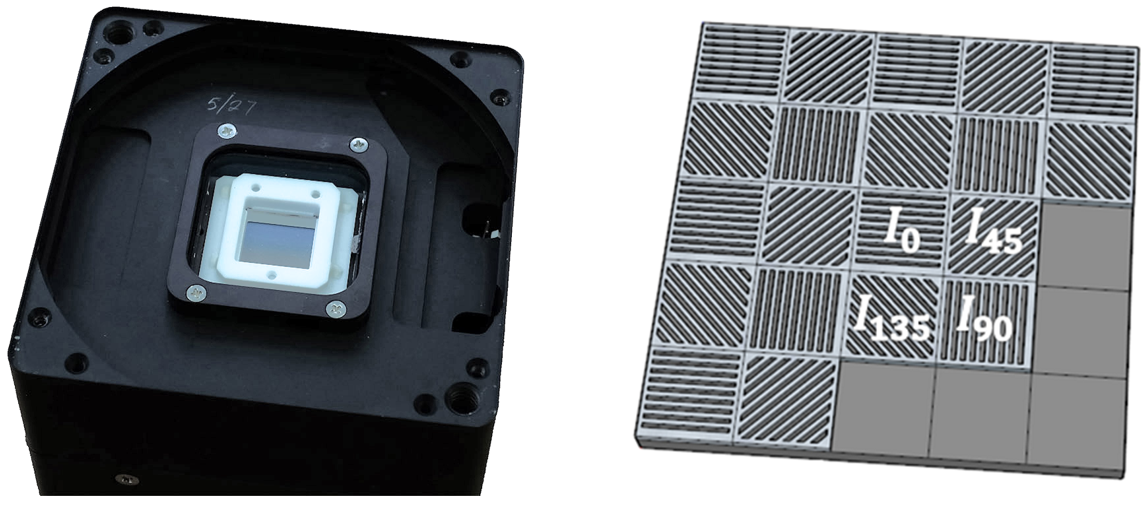}
\caption{\textit{Left:} The RIT Polarization Imaging Camera uses a KAI-04070 interline CCD from On Semiconductor, which is aligned with a micropolarizer array from Moxtek, Inc, by 4D Technology. The polarization sensor is housed in a cooled scientific MicroLine camera from Finger Lakes Instrumentation. \textit{Right:} A schematic representation of the polarization sensor, where each pixel of the imaging sensor is covered by a single wiregrid linear polarizer. The intensities measured by a set of 4 pixels with different angles (0$^{\circ}$, 45$^{\circ}$, 90$^{\circ}$, 135$^{\circ}$) are demodulated to obtain the Stokes parameters. For RITPIC, the micropolarizer array is ahead/above the sensor and microlens array.}
\label{fig:ritpicAndSensor}
\end{figure}

Pixelated polarization sensors represent the most commercially successful and mature implementation of  division-of-focal plane polarimeters (see reviews by \cite{Tyo2006} and \cite{Snik2014}) to date. These devices employ micropolarizer arrays to modulate the intensity of light on a per-pixel basis (Figure \ref{fig:ritpicAndSensor}, Right). Their key advantages are the ability to sample the electric field along several orientations in a single snapshot, compactness, ease of use, and the stability of their calibration \citep{Vorobiev2018}. In 2019, there are two variants of polarization sensors on the market: 1. Commercial off-the-shelf imaging sensors that are aligned and bonded to a micropolarizer array by a third party; 2. CMOS sensors (such as the IMX250MZR) from Sony, which include the micropolarizer array as part of the semiconductor fabrication process, under the microlens array. The RIT Polarization Imaging Camera (RITPIC) used to make the observations in this paper employs the former approach, with the alignment performed by 4D Technology. 

\section{Polarimetry of the Solar Corona} \label{sec:dataAcquisition}

\begin{figure}[ht]
\centering
\includegraphics[width=\textwidth]{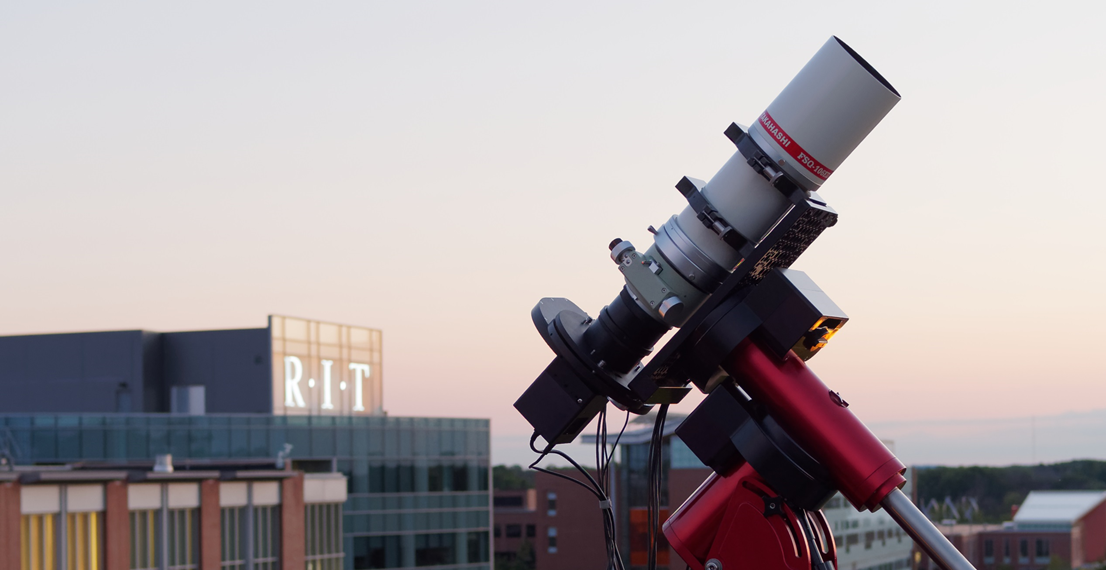}
\caption{The polarimetric system used to observe the total eclipse of 2017 consisted of the RIT Polarization Imaging Camera, filter wheel and focuser from Finger Lakes Instrumentation, and a Takahashi FSQ 530mm f/5 refracting telescope, on an equatorial mount from Software Bisque. Here the system is shown on the roof the Chester F. Carlson Center for Imaging Science at RIT, in Rochester, NY.}
\label{fig:ritpicSystem}
\end{figure}

To perform polarimetry of the Solar corona, we used a \SI{530}{\milli\meter} f/5 Takahashi FSQ 4-element refractive telescope and the RIT Polarization Imaging Camera (RITPIC, \cite{Vorobiev2018}; see Figure \ref{fig:ritpicSystem}). This resulted in a plate scale of \SI{2.88}{\arcsecond} per pixel and \SI{5.76}{\arcsecond} per image element/superpixel. The resulting field of view was $\sim$\SI{63}{\arcminute}$\times$\SI{63}{\arcminute}, \textit{i.e.} roughly 4 R$_{\Sun}$. In this case, the seeing disk is severely undersampled, with only $\sim$0.5 - 1 superpixel per FWHM. This focal length was chosen regardless,because a field of view $\gtrsim$2 R$_{\Sun}$ was needed to study the structure of the corona in the region of 1 - 1.5 R$_{\Sun}$. 

\subsection{Instrumental Characterization}
\begin{figure}[ht]
\centering
\includegraphics[width=\textwidth]{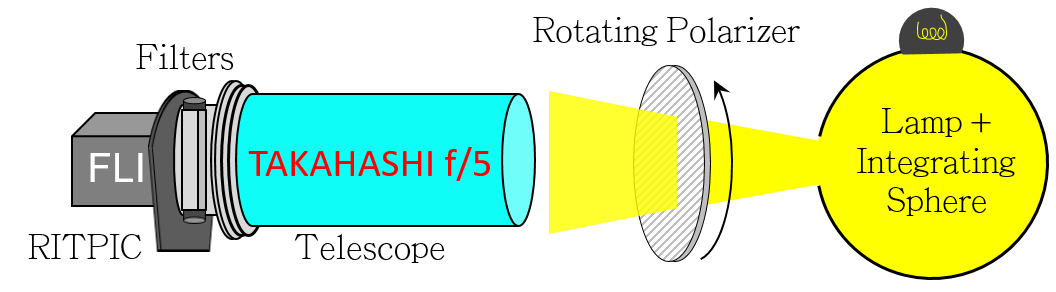}
\caption{The instrumental response of each pixel of the RIT Polarization Imaging Camera was characterized by illuminating the entire system with polarized light, using an integrating sphere and a rotating polarizer. The relative throughput, polarization efficiency, and orientation were determined by model fitting, as described in \cite{Vorobiev2018}.}
\label{fig:characterizationSetup}
\end{figure}

To prepare for observations in the field, the instrumental response of each pixel was determined for the entire system (telescope, filters, and RITPIC) (Figure \ref{fig:characterizationSetup}), using an integrating sphere and a rotating linear polarizer. A thorough discussion of the response non-uniformity of RITPIC and the characterization process is presented in \cite{Vorobiev2018}.

\subsection{Image Acquisition}

Our observations were made from Madras, OR, on August 21, 2017. We acquired sequential images in the Bessel B and R passbands, using a range of exposure times. The intensity of the corona changes by a factor of 200 from the edge of the solar disk to a distance of 1 R$_{\Sun}$ away. Our sensor's limited dynamic range required the use of exposure bracketing (a.k.a. high dynamic range imaging) to image the corona with sufficient SNR, while remaining in the linear operational regime of the CCD. A total of 42 frames (in seven sequences of 6 exposures each) were acquired during totality; a summary of target regions, exposures, and passbands is given in Table \ref{tbl:exposures}. In this way, we obtained a full set of observations at 7 instances over the duration of totality. 

\begin{table}[h]
    \centering
\begin{tabular}{|c|c|c|c|}
\hline  
Filter  & Exposure  & \# of frames & Part of corona   \\
        & time (sec)&              & to be sampled  \\
\hline     
B       & 0.05      & 7            & 1.1-1.25 R$_\odot$  \\
        & 1         & 7            & 1.25-1.5 R$_\odot$   \\
\hline
R       & 0.01      & 7            & 1-1.1 R$_\odot$  \\
        & 0.05      & 7            & 1.1-1.25 R$_\odot$   \\
        & 1         & 7            & 1.25-1.5 R$_\odot$  \\
        & 5         & 7            & 1.5-2 R$_\odot$  \\
\hline
\end{tabular}
    \caption{Exposures acquired during the Aug 21, 2017 total solar eclipse with the RIT Polarization Imaging Camera.}
    \label{tbl:exposures}
\end{table}

\begin{figure*}[ht]
\centering
\includegraphics[width=0.5\textwidth]{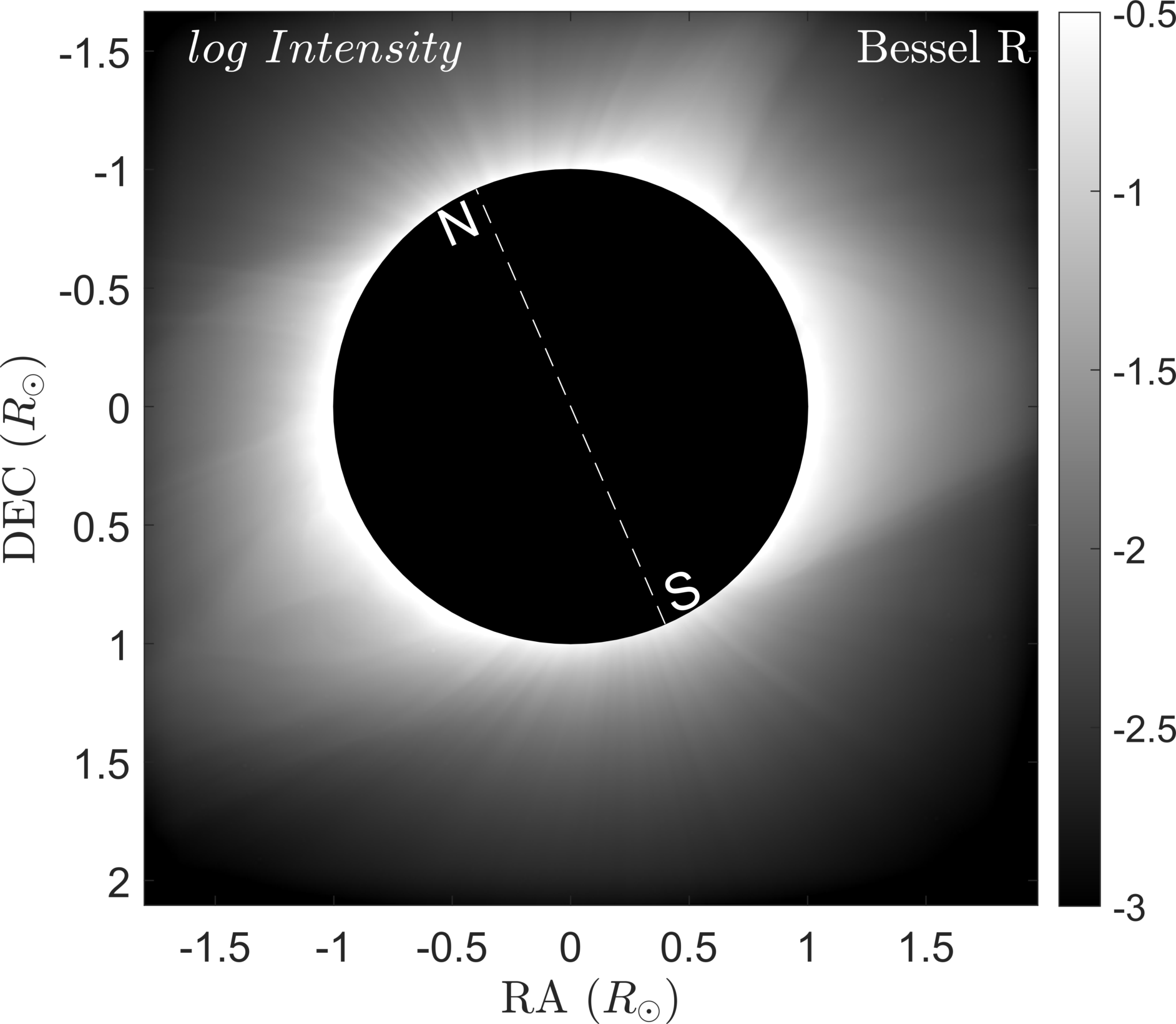}
\caption{This map of the intensity of the corona (Stokes I) was made by median-combining the set of 7 individual Stokes I HDR maps, in the Bessel R band. This set of exposures spans the entire duration of totality.}
\label{fig:coronaIntensity}
\end{figure*}

We acquired observations at four different exposure levels (two for the B filter), with seven images per exposure level,  during totality (see Table \ref{tbl:exposures}). First, maps of the normalized Stokes parameters, $q$ and $u$, were calculated for each of the seven images and combined using a median. This resulted in four median maps, corresponding to the four exposure levels. Then, regions of high (photometric) signal were identified in each map. Pixels with a photometric SNR $<35$ and pixels close to saturation were flagged as ``bad''. The remaining high SNR pixels in the four median maps were combined using a weighted mean; the SNR of each pixel in each map was used as its weight (see Sec. \ref{sec:uncertaintyAnalysis}) . 

\subsection{Polarization of the Solar Corona}
\begin{figure*}[ht]
\centering
\includegraphics[width=\textwidth]{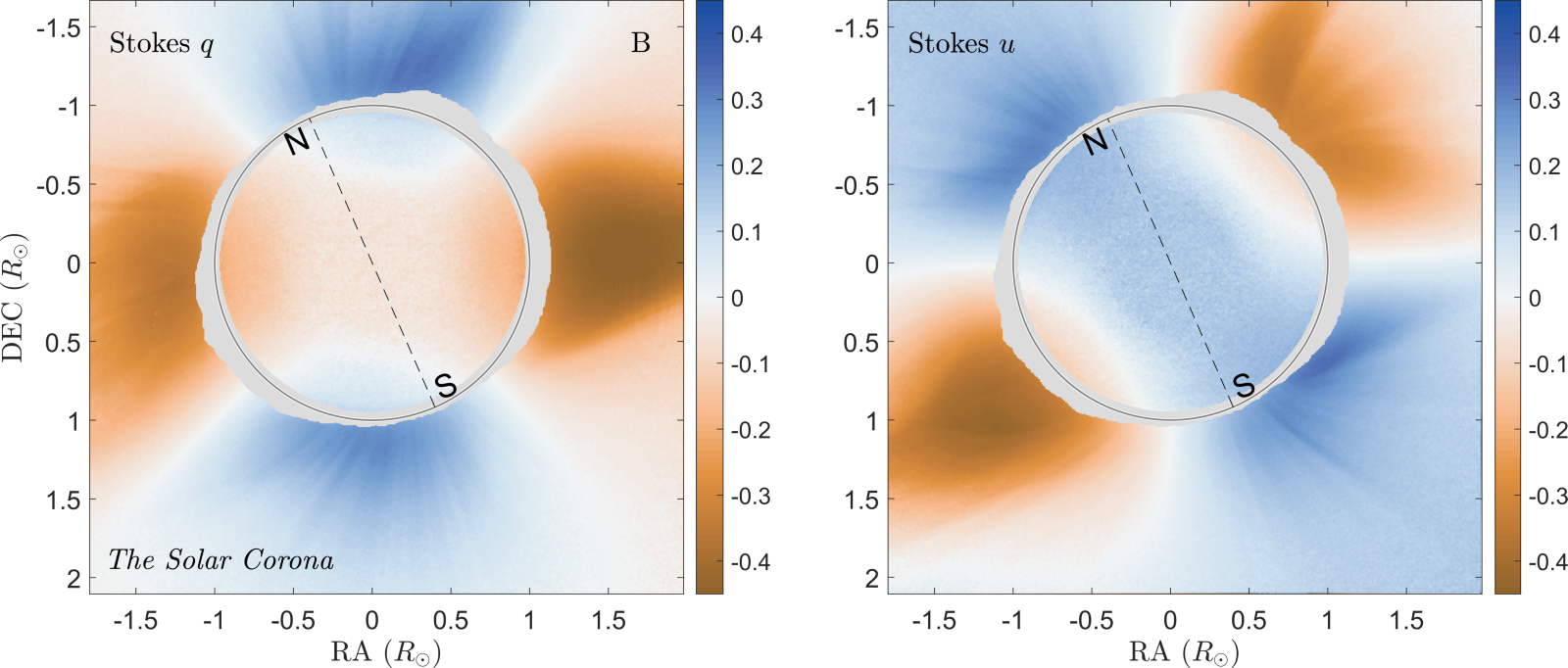}
\includegraphics[width=\textwidth]{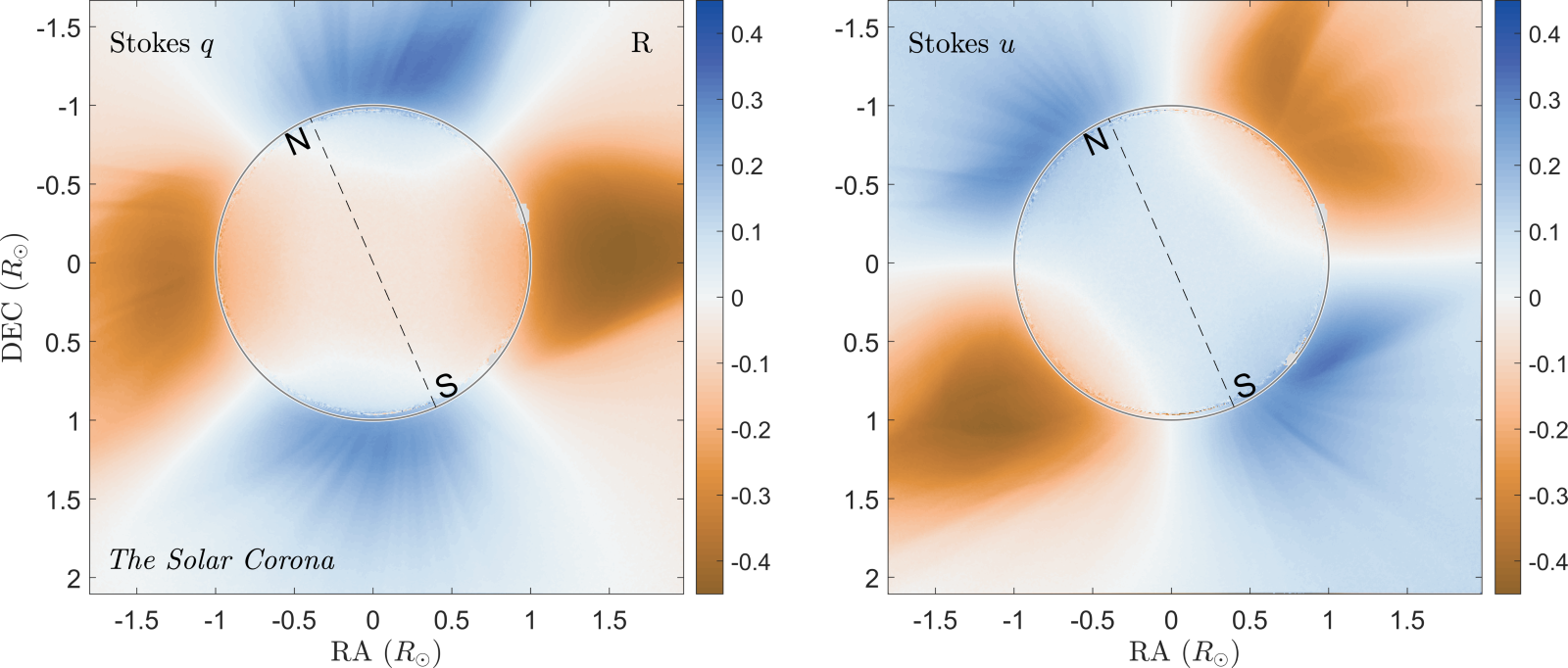}
\caption{Maps of the normalized Stokes parameters, $q$ and $u$ (in the instrument's reference frame), show the polarization structure of the Solar corona in the Bessel B (Top) and Bessel R (Bottom) filter passbands. The general tangential polarization of the corona, as well as some smaller scale features are clearly seen.}
\label{fig:eclipseQU}
\end{figure*}

In this section, we present the results of our linear polarimetry of the Solar corona. First, we show the maps of the normalized Stokes parameters, $q$ and $u$, in the instrumental reference frame. These maps were made by obtaining the median of the 7 HDR frames acquired through the duration of totality (each consisting of 2 exposures for the B channel and 4 exposures for the R channel). The resulting composite maps are shown in Figure \ref{fig:eclipseQU}. The distribution of the normalized Stokes parameters in the corona shows the overall tangential polarization structure, as well as some smaller scale features. The peak fractional polarization is $\sim$47\% in both $q$ and $u$. Overall, the fractional polarization in the B channel appears to be slightly higher than through the R filter.

\clearpage

\begin{figure*}[ht]
\centering
\includegraphics[width=\textwidth]{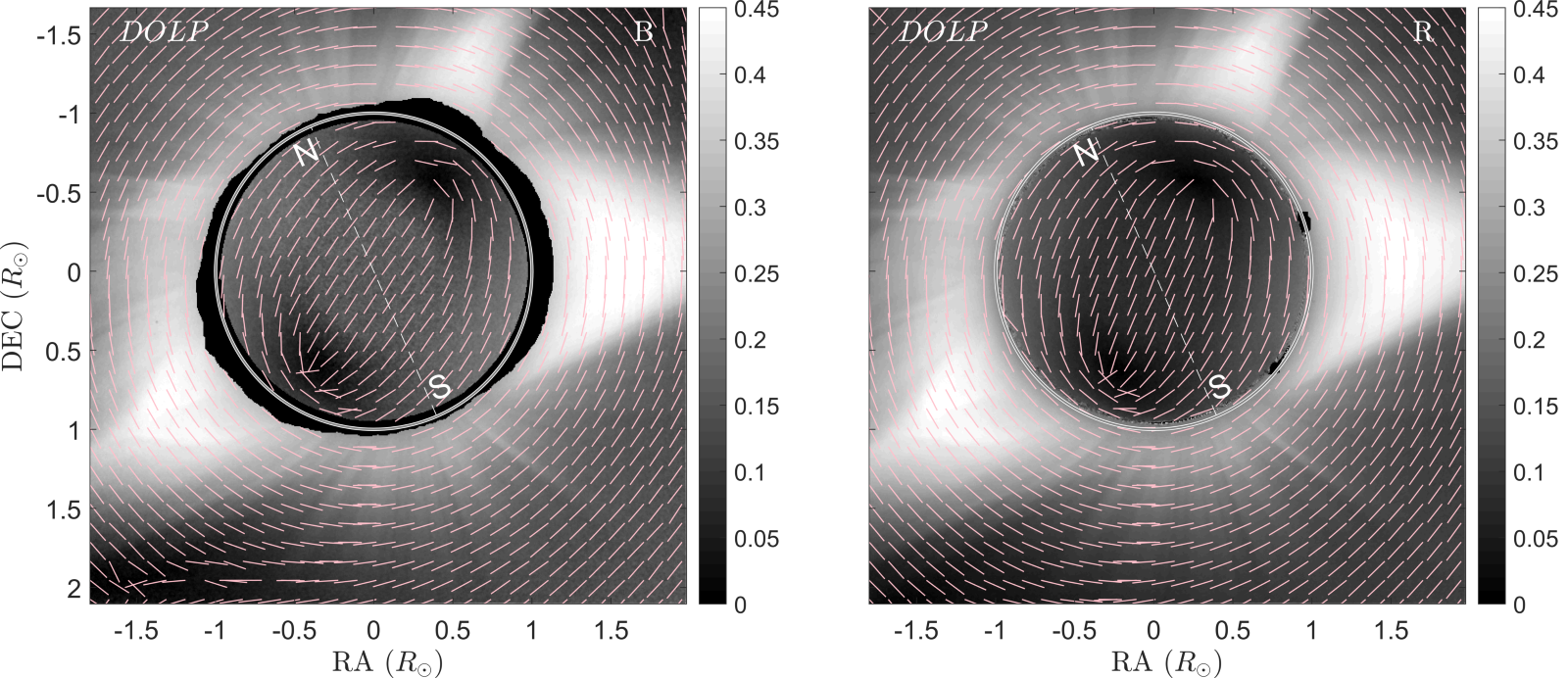}
\caption{The degree of linear polarization (DOLP) and the angle of linear polarization (AOLP) in the instrumental reference frame are shown (for the Bessel B (Left) and Bessel R (Right) filters) as an intensity map and vector plot, respectively.}
\label{fig:dolpAolp}
\end{figure*}

Using the normalized Stokes parameters, we calculate the degree of linear polarization (DOLP) and the angle of linear polarization (AOLP) (Figure \ref{fig:dolpAolp}). Although DOLP is a biased estimator of the fractional polarization, it is useful because it is independent of any reference frame, allowing for straightforward comparison of measurements made by different observers/instruments. 

At a glance, the structure of the DOLP is similar to that of the overall intensity (Figure \ref{fig:coronaIntensity}). However, there exist several differences. Most significantly, the intensity of the corona (Stokes I) peaks near the limb and rapidly decreases with increasing distance from the Sun. Meanwhile, the DOLP is not maximal near the limb ($\sim30\%$), and peaks at a distance of $\sim1.5R_{\Sun}$. Furthermore, the rate at which the intensity and DOLP decrease is dramatically different. Near the edges of our field of view, the intensity has decreased by a factor of $\sim$500, whereas the DOLP has only decreased by a factor of $\sim2$. 

\subsection{Sky Polarization}

\begin{figure*}[ht]
\centering
\includegraphics[width=\textwidth]{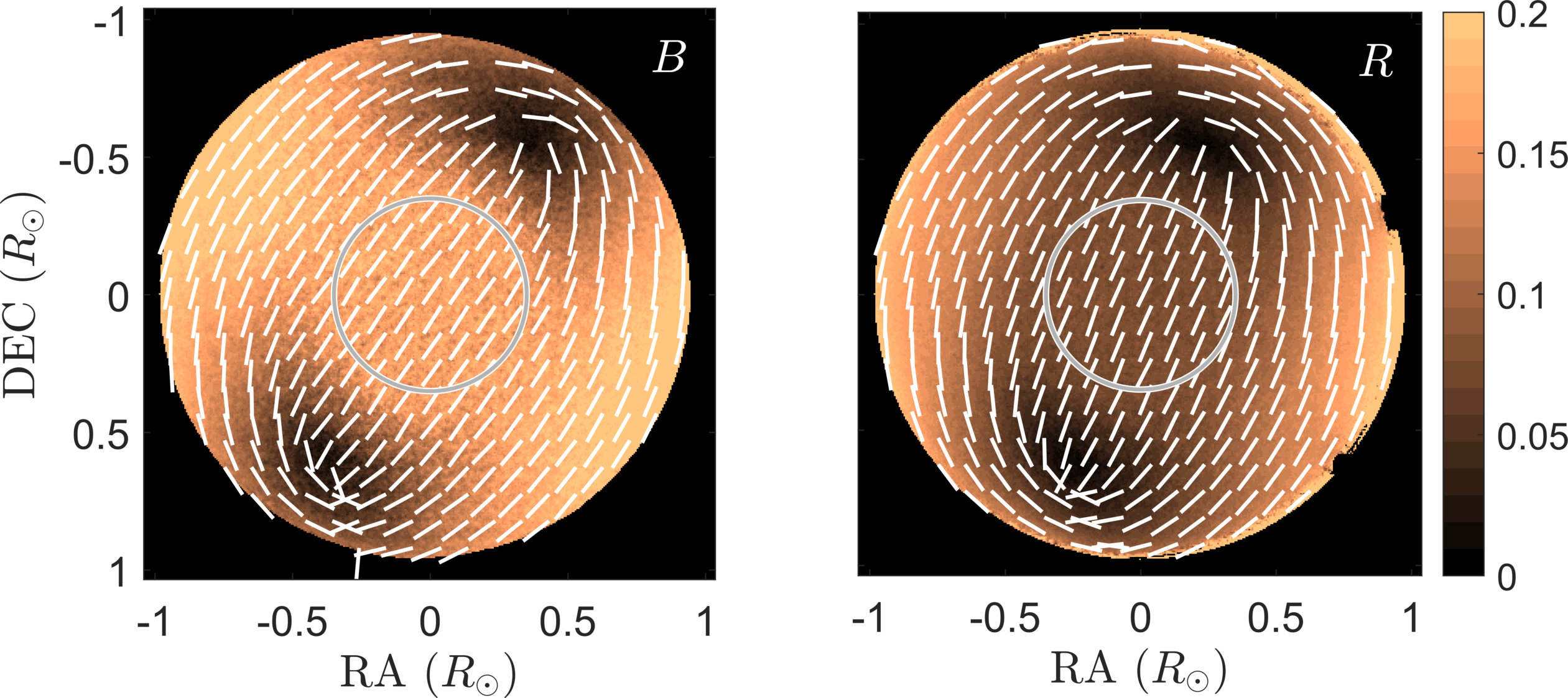}
\caption{We measured the foreground sky polarization in the Bessel B (Left) and R (Right) filter passbands. The gray circle near the center indicates the region we assumed to be dominated by sky foreground and used for foreground subtraction in the following analysis. The median polarization in the B and R bands in this region was 0.16$\pm$0.013 and 0.08$\pm$0.005, respectively.}
\label{fig:skypolarization}
\end{figure*}

As part of an extensive observing campaign prepared for the 2017 eclipse, all-sky polarimetry by \cite{Eshelmann2018} \& \cite{Shaw2019} revealed that the sky in the direction of the sun is polarized in a large band, at the $\sim40\%$ level. This pattern is significantly different than the typical symmetric polarization polarization pattern, which is minimal towards the sun, and increases to a peak at $\sim90^{\circ}$ away from the sun. Our observations have the sufficient spatial resolution and dynamic range to measure the foreground sky polarization across the solar/lunar disk. The degree of polarization for the Bessel B and R channels are shown in Figure \ref{fig:skypolarization}. The linear polarization of the sky is clearly defined up to a distance of $\sim0.5R_{\Sun}$ from the center of the solar/lunar disk, where it begins to become comparable in intensity to the scattered light from the solar corona, which maintains its characteristic circular pattern. The regions where the sky and coronal light have parallel and orthogonal polarization are clearly distinguishable by regions of maximal and minimal fractional polarization, respectively.

The measurements in the blue and red channels exhibit some systematic differences. First, the degree of polarization in the blue channel (as measured in the circled region of Figure \ref{fig:skypolarization}) is higher than in the red channel, at 0.16$\pm$0.013 and 0.08$\pm$0.005, respectively. This is not surprising, as the blue photons scatter more efficiently. However, the angle of polarization of the sky foreground differs between the two channels by $\sim11\pm 2$ deg.

\subsubsection{Sky Foreground Subtraction}
\begin{figure*}[ht]
\centering
\includegraphics[width=\textwidth]{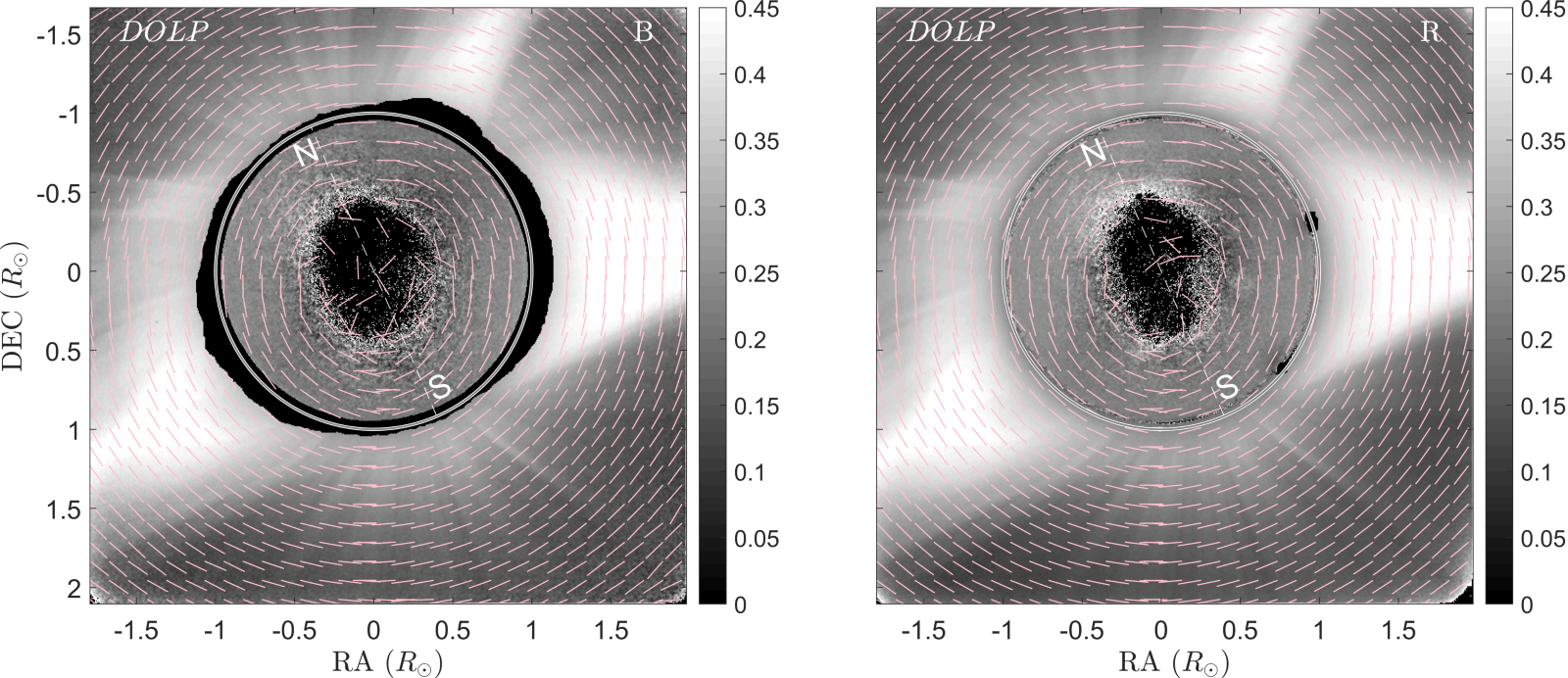}
\caption{The degree and angle of linear polarization for the Bessel B and R channels, after the sky foreground was removed. The scattered light from the corona can be seen across much of the lunar disk. Overall, the angle of linear polarization is more uniform across the field of view, compared to that in Figure \ref{fig:dolpAolp}.}
\label{fig:dolp_aolp_skySubtracted}
\end{figure*}

As part of our polarization analysis, we subtract the sky foreground early in the demodulation process. In each frame, the sky foreground is estimated by the median of the pixels in the region circled in Figure \ref{fig:skypolarization}. We assume that in this region the polarized brightness is dominated by the sky, with minimal contribution from the scattered coronal light. In the discussion that follows, all maps have been foreground-subtracted. The foreground-subtracted polarization maps are shown in Figure \ref{fig:dolp_aolp_skySubtracted}.

 Though the coronal polarization does not change dramatically, some key differences can be seen, as compared to the previous maps (Figure \ref{fig:dolpAolp}). The light behind the lunar disk is now dominated by scattered light from the corona, which maintains its tangential orientation. The DOLP in this region is $\sim23\%$ in the B channel and $\sim25\%$ in the R channel. The angle of linear polarization throughout the field of view is more uniform without the sky foreground, especially at distances greater than $\sim2R_{\Sun}$, as can be seen in the bottom left corner of the Bessel B map. In the future, it would be useful to compare the sky polarization measured behind the lunar disk to that measured at a greater distance from the Sun, for example over the range of $\sim5-15 R_{\Sun}$.   
 
\subsection{Analysis of Uncertainty}
\label{sec:uncertaintyAnalysis}

\begin{figure*}[ht]
\centering
\includegraphics[width=\textwidth]{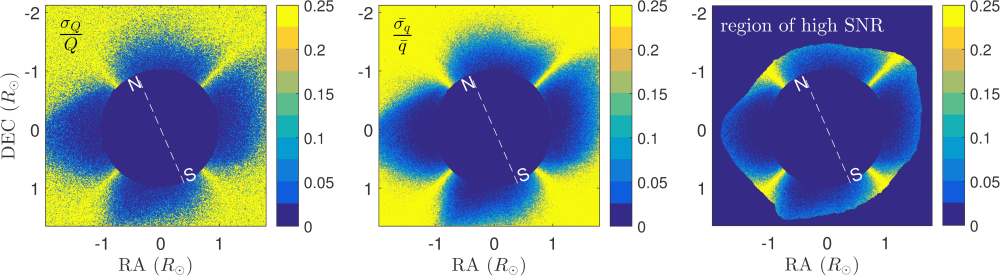}
\caption{\textit{Left:} The fractional error associated with the estimation of Stokes Q in a single \SI{10}{\milli\second} frame. \textit{Middle:} The fractional error of the estimation of Stokes $q$, made from a set of 7 frames. \textit{Right:} A region of the Stokes $q$ map, with photometric SNR $>35$ and excluding saturated pixels.}
\label{fig:errorExamples}
\end{figure*}

In this subsection, we outline the procedure we used to obtain an estimate of the uncertainty associated with our measurements of the normalized Stokes parameters and the DOLP. The Stokes parameters are calculated in each RITPIC frame on a per-pixel basis, using a linear least squares fit and the associated uncertainties are calculated using analysis of residuals. The variance associated with each measurement, $\sigma^2$, is calculated using the unbiased estimator $\hat{\sigma}^2$ \citep{Bajorski2011} as follows,

\begin{equation}
\label{eq:varianceEstimator}
\hat{\sigma}^2=\frac{1}{n-2}\sum_{i=1}^{n}(y_i - \hat{y}_i)^2  = \frac{1}{2}\sum_{i=1}^{n}(y_i - \hat{y}_i)^2,
\end{equation}

where $y_i$ is an observed intensity and $\hat{y}_i$ is the model fit. This results in a map of variance for each Stokes parameter ($\hat{\sigma}_I$, $\hat{\sigma}_Q$, $\hat{\sigma}_U$,), for each data frame (Figure \ref{fig:errorExamples}, \textit{Left}). Using formal uncertainty propagation, we determine the variances of the normalized Stokes parameters: $\sigma_q$ and $\sigma_u$. Next, using the set of 7 measurements, we obtain the median values of the normalized Stokes parameters and the average variance:  $\bar{q}$ and $\bar{\sigma^2_q}$ (Figure \ref{fig:errorExamples}, \textit{Middle}). Finally, we use the map of Stokes I to find regions with photometric SNR $>$ 35 and mask off saturated pixels. The resulting region of ``high quality'' data for the \SI{10}{\milli\second} exposures is shown in the right panel of Figure \ref{fig:errorExamples}.

This process is repeated for the remaining frames, with exposures of \SI{50}{\milli\second}, \SI{1}{\second}, and \SI{5}{\second}. The high SNR regions from each exposure set are combined to generate the maps shown in Figure \ref{fig:eclipseQU} and Figure \ref{fig:dolpAolp}. The uncertainties from high SNR regions of each exposure set are combined using a weighted mean. The weighting is determined by the SNR of each measurement. In this way, the contribution of the noisier measurements is diminished, if better data are available from a different exposure set. As a result, the uncertainties of the final measurements, $\sigma_q$ and $\sigma_u$, are not the true standard deviation; however, they are still a reasonable estimate of the uncertainty of each measurement. 

\begin{figure*}[ht]
\centering
\includegraphics[width=\textwidth]{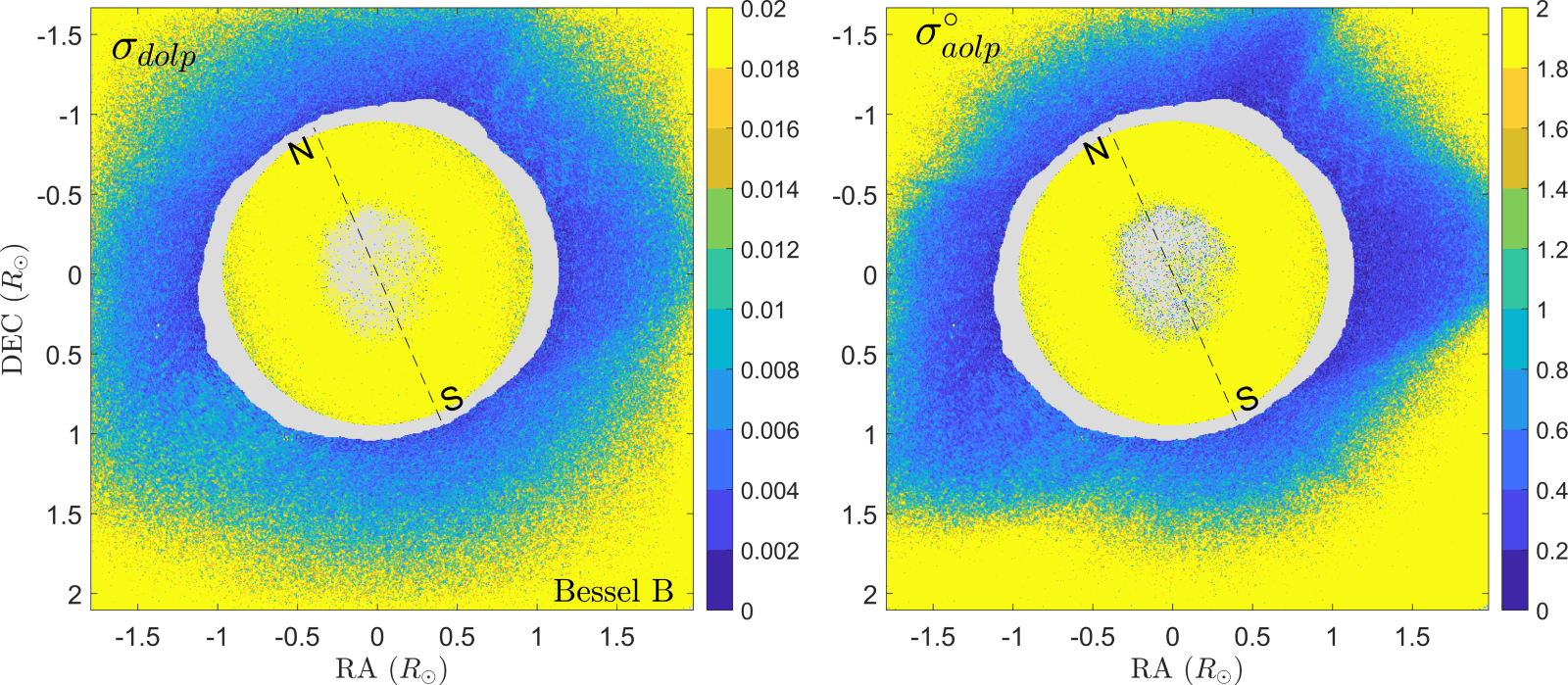}
\includegraphics[width=\textwidth]{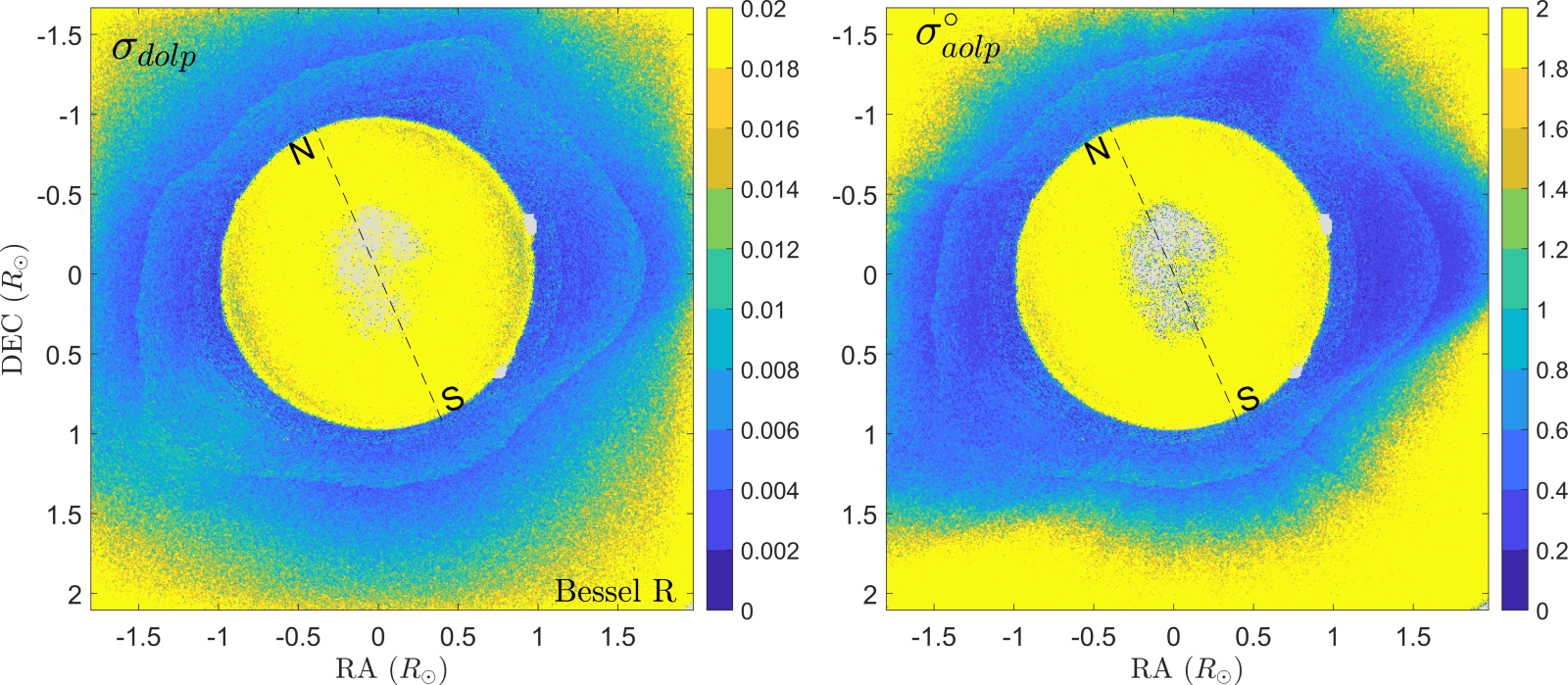}

\caption{\textit{Left:} The uncertainty of the estimation of degree of linear polarization of the corona, estimated as the standard deviation of a set of 7 HDR measurements. In regions of high polarization, the standard deviation is $<1\%$, increasing to $\sim2\%$ near the edges. \textit{Right:} The uncertainty (in degrees) of the angle of linear polarization in our measurement. Within $\sim1.5 R_{\Sun}$, our angle measurement is precise to $\sim$\SI{1}{\degree}. Grey pixels indicate areas of saturation or SNR $<$ 1.}
\label{fig:dolp_aolp_errors}
\end{figure*}

Lastly, we use the uncertainties $\sigma_q$ and $\sigma_u$ to find the uncertainty associated with the degree of linear polarization and angle of linear polarization in the corona, $\sigma_{dolp}$ and $\sigma_{aolp}$, using formal (linear) uncertainty propagation. Strictly speaking, this procedure gives a biased estimate of both the fractional polarization, DOLP, and its uncertainty, $\sigma_{dolp}$. However, the corona is very strongly polarized, even in areas of low polarization. This high ``polarimetric SNR'' ($>$20 across our field of view) ensures that the relative effect of the bias is small. The maps of uncertainty are shown in Figure \ref{fig:dolp_aolp_errors}. In regions where the polarization is large, our combined measurement of the fractional polarization has an uncertainty of $\sim1\%-3\%$. At distances $>1.5 R_{\Sun}$, the lack of signal from the corona and the sky foreground lead to rapidly increasing uncertainty. A similar trend can be seen in the estimate of the angle of linear polarization (Figure \ref{fig:dolp_aolp_errors}, \textit{Right}). Where the signal is high, the angle estimation is precise to $\sim$\SI{1}{\degree}. Further out, the uncertainty increases to $\sim$\SI{2}{\degree}. 

\clearpage
\subsection{The Structure of the Corona}
During this eclipse, the corona showed strong polarization features in the equatorial direction, with weaker polarization (and intensity) at the poles. Large scale and small scale ``streamers'' are clearly visible. The degree of linear polarization of the corona peaks at $\sim47\%$ in the equatorial features, at a distance of $\sim1.5R_{\Sun}$. 

\begin{figure*}[ht]
\centering
\includegraphics[width=0.5\textwidth]{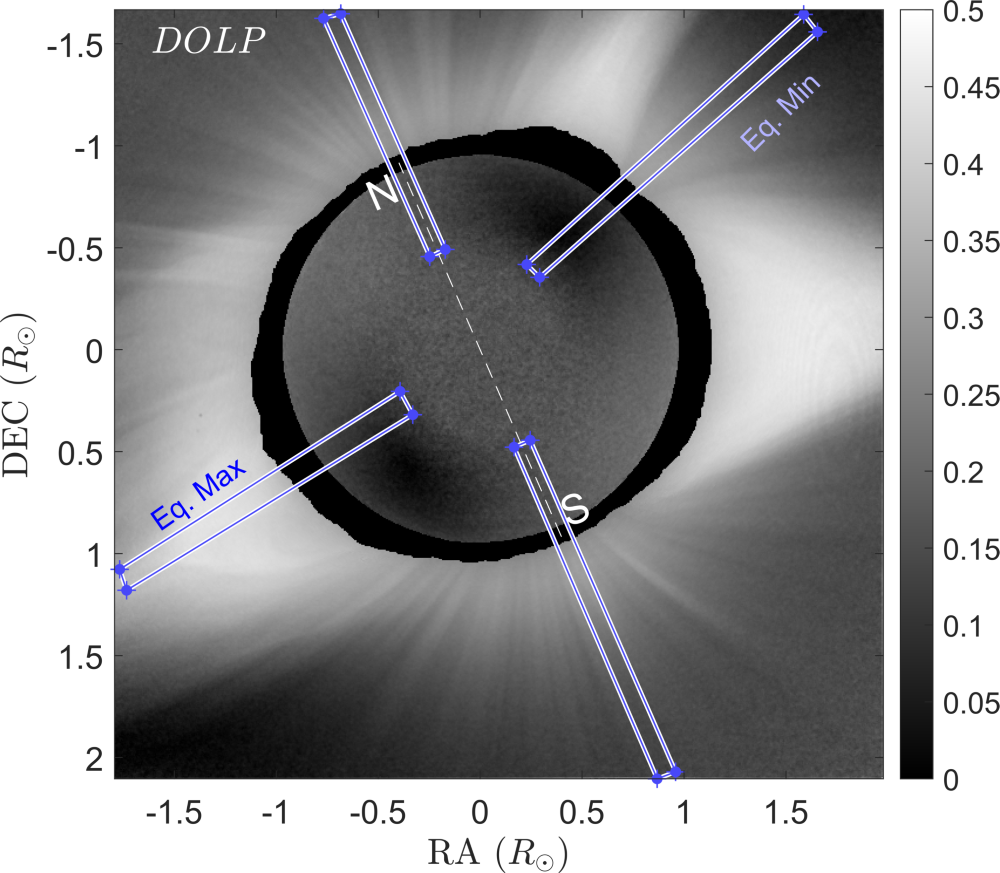}
\caption{To compare our measurements to the predictions of the model corona of \cite{vanDeHulst1950}, we acquired azimuthally-averaged radial profiles in four regions: at the north and south poles, and in regions of maximal and minimal polarization near the equator.}
\label{fig:dolpProfileRegions}
\end{figure*}

For a quick comparison of our measurements to the predictions of a Thomson scattering corona, we use the theoretical curves of \cite{vanDeHulst1950}, which take into account the combination of both the K and F corona. The radial profiles were created by averaging the polarization in the rectangular regions shown in Figure \ref{fig:dolpProfileRegions}, in the azimuthal direction. Four regions were chosen: one each at the north and south poles, and in areas of maximal and minimal polarization in the equatorial regions. 

The maximal polarization we measured agrees well with the maximum predicted values from \cite{vanDeHulst1950}. However, the polarization measured at the north and south poles is significantly higher than the model prediction. The minimally polarized regions near the equator are also more significantly polarized. The source of the discrepancy is not clear to us at this time. It is possible that the spherically symmetric model does not adequately account for the electron density structure in the real corona. Also, the impact of the sky foreground is difficult to rigorously quantify with the data we have available. Indeed, the sky-subtracted maps show more significant deviations from the model than the un-calibrated measurements, especially further from the Solar center. This may indicate over-subtraction of the sky foreground, because the corona intensity (and thus polarized intensity) is fainter in these regions.  

\begin{figure*}[ht]
\centering
\includegraphics[width=\textwidth]{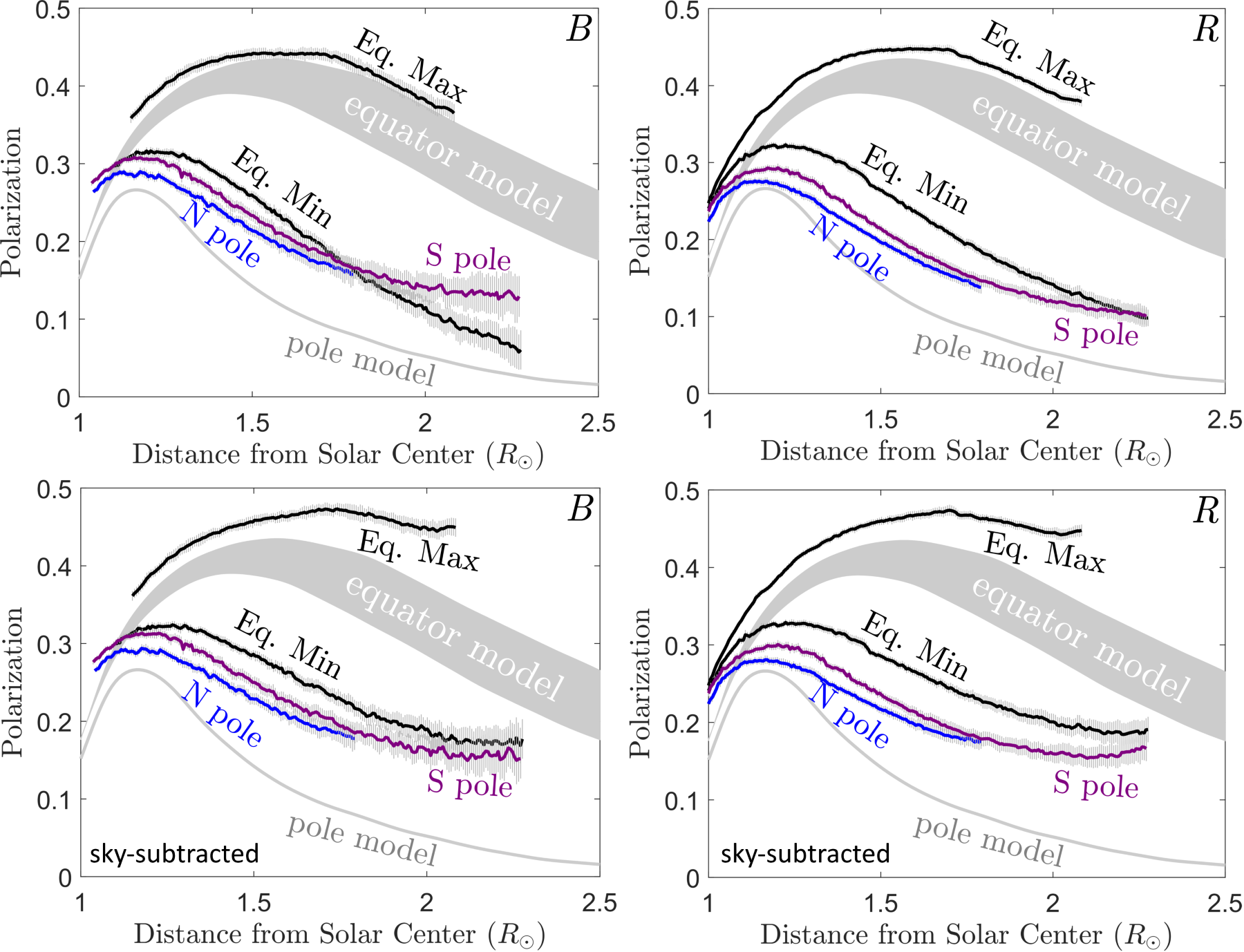}
\caption{Our measured profiles calculated in the four regions indicated in Figure \ref{fig:dolpProfileRegions}, plotted on top of predictions for a Thomson scattering corona from \cite{vanDeHulst1950}. The maximum measured polarization does not greatly exceed the maximum value
predicted by Thomson scattering, up to a distance of $\sim1.75 R_{\Sun}$. However, in profiles where the sky-subtraction was performed (Bottom Row), start to deviate more significantly from the \cite{vanDeHulst1950} model at distances $>1.5 R_{\Sun}$.}
\label{fig:dolp_profiles}
\end{figure*}

\subsubsection{Distribution of the Polarization Angle}
\begin{figure}[ht]
\centering
\includegraphics[width=\textwidth]{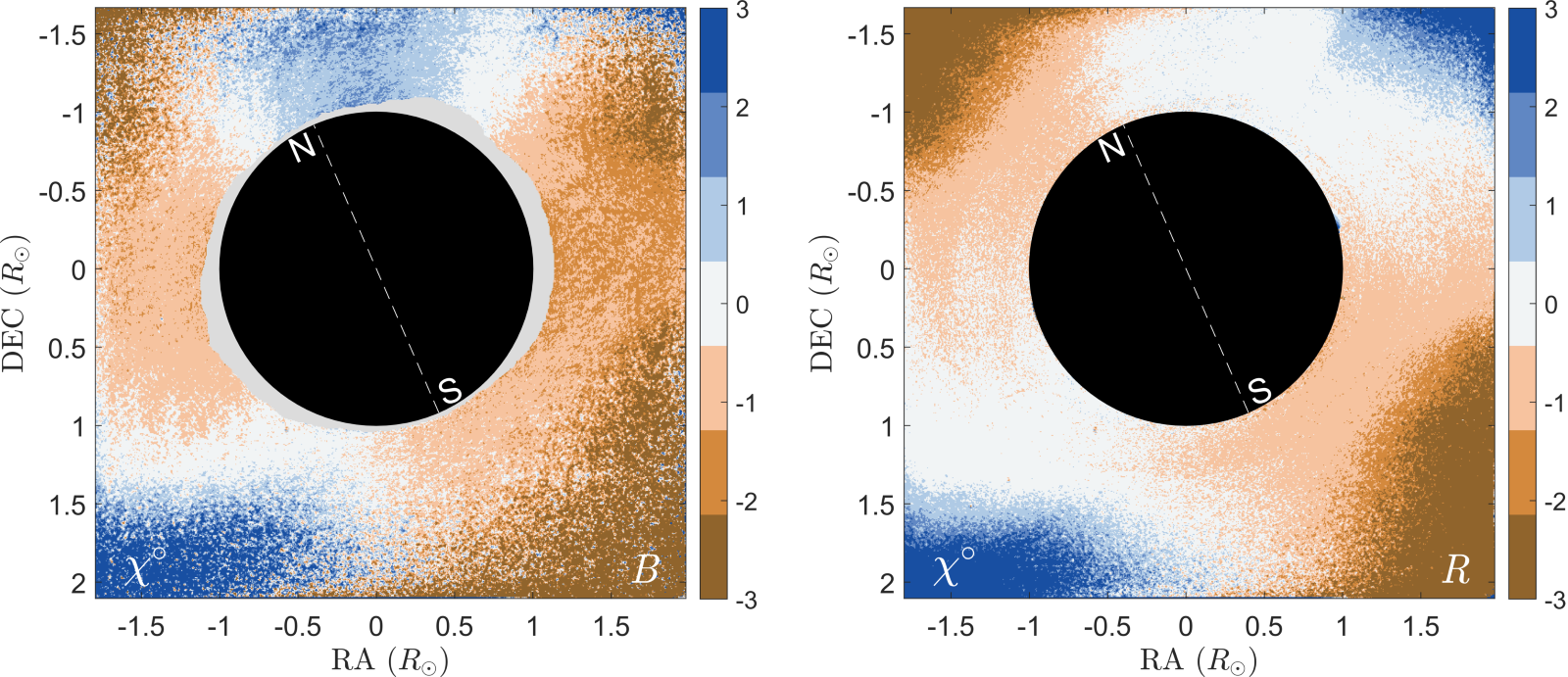}
\caption{A map of the deviation of the angle of linear polarization from the tangential direction, in the B and R filters. Across our field of view, the AOLP follows the tangential direction, within our measurement uncertainties. The grey annulus in the Bessel B map indicates regions of saturation, in our shortest exposures.}
\label{fig:angleX}
\end{figure}

The angle of polarization is tangential to the solar limb.  This is shown in Figure \ref{fig:corona_dolp_aolp}. To identify any underlying structure in the angle of linear polarization, we plot the deviations of the AOLP from the purely tangential direction, denoted by the angle $\chi$, in Figure \ref{fig:angleX}. In regions of strong polarization (and high SNR), our estimate of the angle of linear polarization is consistent with the tangential direction within $\sim$\SI{0.5}{\degree}. Where the polarization is low, like near the south east limb, the deviation increases to $\sim$\SI{1}{\degree}. In regions further than $\sim1.5 R_{\Sun}$, the deviation increases to $>$\SI{3}{\degree}.  
 
\section{Polarization of the Solar Corona}
We performed broadband imaging polarimetry of the corona in the Bessel B and R filters. During the eclipse of August 21, 2017 the corona was asymmetric, with a pronounced ``Y'' shape in the equatorial regions. The peak polarization was $47\%\pm1\%$, in the arms of the ``Y'' structure, at a distance of $\sim1.5 R_{\Sun}$. We do not measure any unexpectedly large polarization, of the kind discussed by \cite{Koutchmy1971} and reported more recently by \cite{Qu2013}. 

Deviations of the angle of polarization from the tangential direction have been reported by various groups, most recently by \cite{Skomorovsky2012}, \cite{Qu2013}, and \cite{Kim2017}. We do not measure any deviations from the tangential direction, larger than \SI{1}{\degree}. The deviation, $\chi$ increases in areas of low SNR, at distances $\gtrsim 2 R_{\Sun}$. We suspect that in these regions, the measurements are becoming affected by the sky polarization. Indeed, \cite{Skomorovsky2012} admit that the strong deviations they measured are ``apparently caused by the insufficient quality of [the] coronal images.'' 

Our measurements are similar to those of \cite{Koutchmy1993}, who did not measure deviations larger than \SI{1}{\degree}. Small deviations of \SI{1}{\degree} - \SI{2}{\degree} have been proposed to arise from scattering by relativistic electrons by \cite{Molodensky1973}, and further developed by \cite{Inhester2015}. The precision of our angle estimation should be sufficient to detect changes of this magnitude, because the beamed emission also shows larger than average fractional polarization, leading to intrinsically higher polarimetric SNR. However, it's possible that our spatial resolution of \SI{5.76}{\arcsecond} ``averages out'' these features, which may only exist on small scales. In any case, it appears that our measurements are sampling the ``thermal noise of the corona'' \citep{Inhester2015}, with no evidence of any other structure. 

\section{Conclusions} Observations obtained with the RIT Polarization Imaging Camera of the 2017 total Solar eclipse represent some of the best measurements of this sort to date. Polarization sensors appear well suited to these observing campaigns, which often involve travel to remote sites and a high degree of automation. The snapshot capability, especially, has proven its utility during this campaign, due to the presence of low, thin, fast moving clouds above Madras, Oregon.

Nevertheless, some challenges remain to be solved. Our measurements cannot easily be used to calculate the polarized brightness, \textit{pB}, in absolute radiometric units (W m$^{-2}$ sec$^{-1}$ nm$^{-1}$), because we cannot easily determine the atmospheric transmission, at the time of totality. Going forward, it may be useful to observe a star of known brightness, in the vicinity of the Sun, to calibrate the transmission of the atmosphere. Similarly, contemporaneous measurements over a larger field of view may be useful to better measure the sky foreground near the corona. 

\section{Acknowledgements}
The authors acknowledge the support and funding provided by Moxtek, Inc. In particular, we thank Ray West and Roger Ketcheson. We thank Dr. Peter Zimmer for providing access to the observing site in Madras, Oregon and Dr. Billy Vazquez for supplying the telescope mount used for this work. 

\bibliographystyle{aasjournal}
\bibliography{Bibliography}{}



\end{document}